# MAGNETOELECTRIC EFFECT FOR NOVEL MICROWAVE DEVICE APPLICATIONS

*E.O. Kamenetskii, M. Sigalov, and R. Shavit*

Microwave Magnetic Laboratory
Department of Electrical and Computer Engineering
Ben Gurion University of the Negev, Israel
*e-mail: kmntsk@ee.bgu.ac.il*

**ABSTRACT**

In a normally magnetized thin-film ferrite disk with magnetic-dipolar modes, one can observe magnetoelectric oscillations. Such magnetoelectric properties of an electrically small sample can be considered as very attractive phenomena in microwaves. In this paper we discuss the question on novel microwave device applications based on ferrite magnetoelectric particles.

*Index Terms* — Ferromagnetic resonance, Magnetoelectric effect, Microwave devices

## 1. INTRODUCTION

An idea to create microwave devices where the magnetic characteristics are controlled by an electric field and/or the electric characteristics are controlled by a magnetic field – the magnetoelectric (ME) devices – looks as a very attractive subject for novel applications. Presently, ME interaction in ferrite-ferroelectric composites have facilitated a new class of microwave signal processing devices. When a ferrite in a ferrite-ferroelectric bilayer is driven to the ferromagnetic resonance (FMR) and an electric field is applied across a ferroelectric, the ME effect results in a frequency or field shift of FMR. Thus microwave devices based on FMR can be tuned with both electric and magnetic fields. Several dual tunable microwave ferrite-ferroelectric-bilayer devices have been demonstrated so far. There are microwave attenuators, resonators, filters, and phase shifters [1].

The ability to control magnetism with electric fields has an obvious technological appeal. Together with the way of development and creation of ME devices based on ferrite-ferroelectric bilayers, unique microwave devices can be realized based on so-called ferrite ME particles. The ME response requires simultaneous breaking of space inversion and time-reversal symmetries. The electric polarization is parity-odd and time-reversal-even. At the same time, the magnetization is parity-even and time-reversal-odd. One cannot consider (classical electrodynamically) a system of two coupled electric and magnetic dipoles as local sources of the ME field [2]. These symmetry relationships make questionable an idea of a simple combination of two (electric and magnetic) small dipoles to realize local ME particles for electromagnetics. In a presupposition that a particle with the near-field cross-polarization effect is really created, one has to show that inside this particle there are internal dynamical motion processes with special symmetry properties. It has been found that while the ME properties are not allowed in classical point-like particles, small ferrite disks with magnetic eigen oscillations, magnetic vortex and chiral structures may present a solution [3]. Recent studies, both theoretical [4] and experimental [5], show unique ME properties of a quasi-2D ferrite disk with magnetic-dipolar-mode oscillations. Small ferrite ME particles demonstrate strong resonance responses to RF electric and magnetic fields in a local region – the region with sizes much less than the free-space electromagnetic wavelength. The near fields of ferrite ME particles are characterized by special symmetry properties. This allows considering ferrite ME particles among the most promising structures for novel microwave applications.

## 2. FERRITE ME PARTICLES

In a normally magnetized ferrite disk with magnetic-dipolar-mode (MDM) [or magnetostatic-wave (MSW)] oscillations, one can observe the ME properties in microwaves. For the first time, the MDM (or MSW) oscillations were found in small ferrite spheres [6]. Further experiments with thin ferrite disks [7] showed very rich multiresonance MDM spectra in such particles. In 1970s – 1980s, the main stream in studies of microwave devices based on magnetostatic waves and oscillations was aimed to realization of compact delay lines, filters, and planar resonators [8]. A strong interest in this kind of microwave devices was caused by the fact that magnetostatic waves have a very small wavelength (two-four orders of a magnitude less than the electromagnetic-wave wavelength at the same frequency). A return to studies of the MDM spectra in thin-film ferrite disks was made recently. It was shown that MDMs in a normally magnetized ferrite disk are energy-eigenstate oscillations. Moreover it has been proven that the MDM orthogonality relations can be satisfied when there are effective surface magnetic currents on a lateral surface of a disk. Due to these currents one has eigen electric fluxes and eigen electric (anapole) moments. For every MDM, there exists also a vortex circulation of the power flow density inside a ferrite disk. Unique symmetry properties of MDMs in a ferrite disk result in appearance of microwave ME effects [4]. These effects were verified in different microwave experiments [5].

Recently, it was shown [9] that ferrite ME particles can be effectively studied based on the HFSS numerical simulation program [10]. Fig. 1 shows a typical picture of the vortex circulation of the power flow density inside a ferrite disk for the 1st MDM.

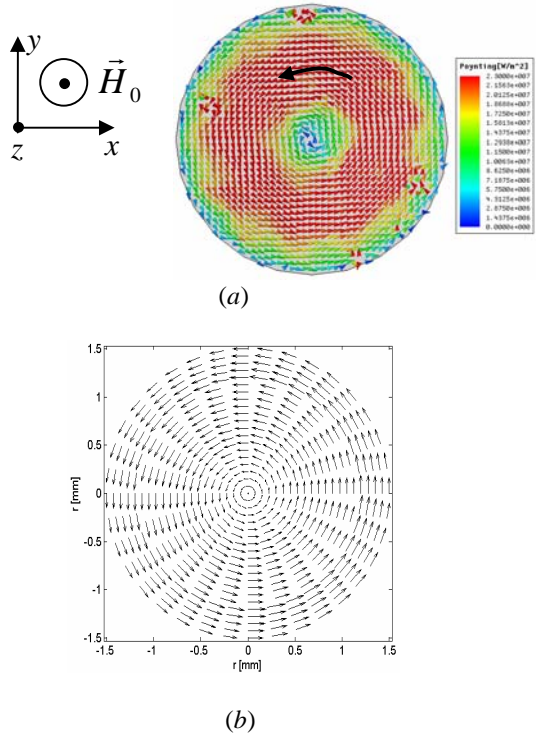

Fig. 1. The power flow density distribution for the 1st MDM in a quasi-2D ferrite disk. The frequency is $f = 8.52$ GHz and the bias magnetic field $H_0 = 4900$ Oe. The disk diameter is $D = 3$ mm and the thickness is $t = 0.05$ mm. (a) Numerically modeled vortex; (b) analytically derived MDM vortex. A black arrow in Fig. 1 (a) clarifies the power-flow direction inside a disk.

As an example of ME properties, Fig. 2 shows numerical results of the electric field distributions at different time phases for the 1st oscillating mode in a ME particle composed as a thin ferrite disk with a wire surface metallization. One has an in-plane non-rotating electric dipole. At the same time a magnetic dipole rotates in a disk plane. The concept of such a ferrite ME particle was proposed in [11].

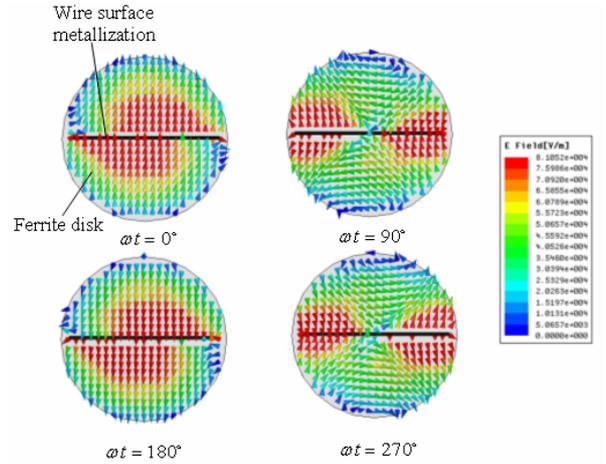

Fig. 2. The electric field structure in a ferrite ME particle composed as a thin ferrite disk with a wire surface metallization.

It appears now that the MDM ME particles can be considered among the most promising structures for novel microwave applications. There could be: (a) high-resolution near-field microwave sensors, (b) dense microwave ME metamaterials, (c) novel compact microwave radiating systems, and (d) microwave electrostatic-control spin-based logic devices and quantum-type computation.

## 3. ME NEAR-FIELD MICROWAVE SENSORS

Measurement of the electromagnetic response of materials at microwave frequencies is important for both fundamental and practical reasons. During last years near-field sensors for microwave microscopy have created the opportunity for a new class of electrodynamics experiments of materials and integrated circuits [12]. One of the most sensitive forms of contemporary near-field microwave microscopy is that the sample is put near the open end of a transmission-line resonator, and changes in the resonant frequency and quality factor are monitored as the sample is scanned. It becomes clear that new perfect lenses that can focus beyond the diffraction limit could revolutionize near-field microscopy. We propose use of a small ferrite ME particle as an effective near-field sensor for novel microwave microscopes.

Due to special symmetry properties of magnetic ordering in thin ferrite disks with MDM oscillations there exist eigen electric fluxes. These fluxes should be very sensitive to the permittivity parameters of materials abutting to the ferrite disk. Dielectric samples above a ferrite disk with a higher permittivity than air confine the electric field closely outside the ferrite, thereby changing the loop magnetic currents in a ferrite disk and thus transforming the MDM oscillating spectrum. This effect was verified experimentally and explained theoretically in Refs. [4, 5]. Fig. 3 shows the experimental results of the frequency shift of the MDM peak positions for a ferrite disk with a dielectric loading due to the eigen electric fluxes. The first peaks in the spectra are matched by proper correlations of bias magnetic fields.

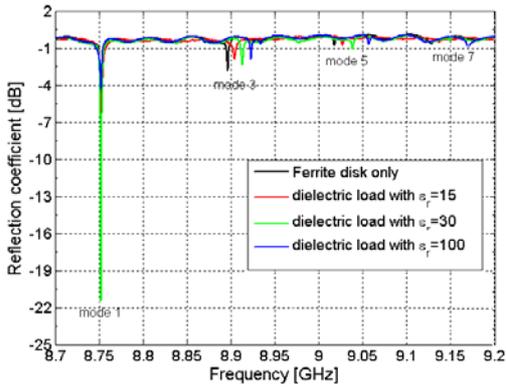

Fig. 3. Transformation of the MDM oscillating spectrum for different dielectric loads.

Based on ferrite ME particles, effective near-field sensors for novel microwave microscopes can be realized. The main principle of the near-field microscopy is the use of evanescent modes decaying exponentially. These modes are superoscillatory and thus provide a means to probe high spatial frequency structure of the sample. The scattered field is computed perturbatively. Since the wavelength of MDMs is three orders of magnitude less than the electromagnetic-wave wavelength, the ferrite-particle sensors should have very high-resolution characteristics.

### 4. DENSE MICROWAVE ME METAMATERIALS

Presently, there is a strong interest in electromagnetic artificial materials with local ME properties. The properties of ferrite ME particles and ME fields originated from such particles give very new insights into a problem of dense microwave ME composites. Such materials can be useful for novel microwave waveguides and antenna systems.

In 1948 Tellegen suggested that an assembly of the lined up electric-magnetic dipole twins can construct a new type of an electromagnetic material [13]. Till now, however, the problem of creation of the Tellegen medium is a subject of strong discussions. An elementary analysis makes questionable an idea of a simple combination of two (electric and magnetic) dipoles to realize local materials with the Tellegen particles as structural elements. A classical multipole theory describes an effect of "ME coupling" when there is time retardation between the points of the finite-region charge and current distributions and this time retardation is comparable with time retardation between the origin and observation points. In such a case, an expression for the field contains combinations of both magnetic and electric multipole moments. One may obtain the EM-wave phase shift between the points of the finite-region charge and current distributions, $\varphi_1$, comparable with the EM-wave phase shift between the origin and observation points, $\varphi_0$, even for a very small scatterer. To obtain such an effect of "ME coupling" one should make a scatterer in a form of a small $LC$ delay-line section. In the far zone of this scatterer we will observe "ME coupling". This can be explained with help of Fig. 4. Let a characteristic size of a scatterer be $r$ and $r << R$, where $R$ is a distance between the origin point and the observation point $P$. Let $k_1$ be the wavenumber of the EM wave propagating in a $LC$ delay line and $k_0$ be the wavenumber of the EM wave in vacuum. In a case when $k_1 >> k_0$, one may obtain $\varphi_1 = kr \approx \varphi_0 = k_0 R$. All the proposed "electromagnetic ME scatterers" have a typical form of a delay-line section with distinctive inductive and capacitive regions. In a series of experimental papers one can see that the "ME coupling" effect in these particles was observed only in the propagation-wave behavior, without any near-field characterizations.

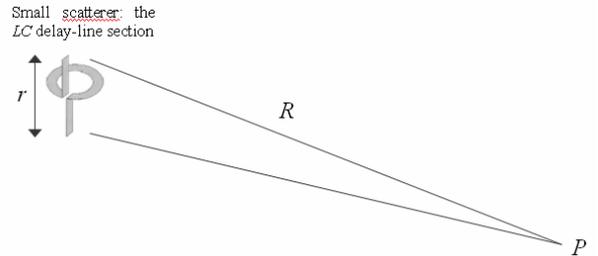

Fig. 4. Effect of "ME coupling" in a small electromagnetic scatterer. A scatterer has a form of a $LC$ delay-line section and the phase shifts are $\varphi_1 = kr \approx \varphi_0 = k_0 R$

To realize dense microwave ME materials, local ME structural elements should be used. While a local ME particle cannot be realized as a classical scatterer with the induced parameters, it can be created as a small magnetic sample with eigen magnetic oscillations having special symmetry breaking properties [3]. A model for coupled ferrite ME particles underlies a theory of ME "molecules" and dense ME composites. The model is based on the spectral characteristics of MDM oscillations and an analysis of the overlap integrals for interacting eigen oscillating ME elements [14]. Fig. 5 gives the numerical results of the magnetic-field distributions for even and odd modes in two coupled ferrite particles.

The chiral-state resonances in a ferrite disk can be described by helical MS-potential modes [15]. Because of the symmetry breaking effects, dense ferrite-particle ME composites will have the left-hand-resonance and the right-hand-resonance responses. Fig. 6 represents a dense microwave ME composite material with an illustration of helical (chiral) states of the MDMs inside ferrite particles.

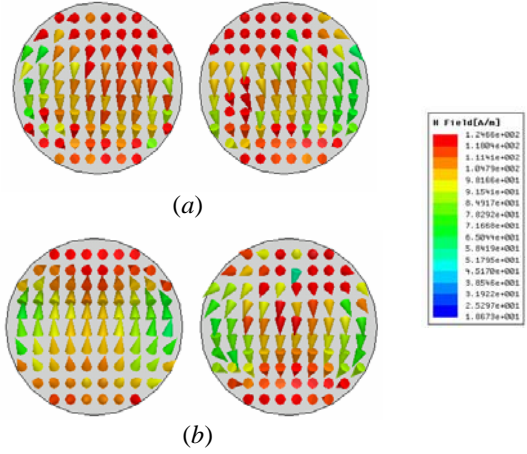

Fig. 5. The magnetic-field distributions for even and odd modes in two coupled ferrite particles.

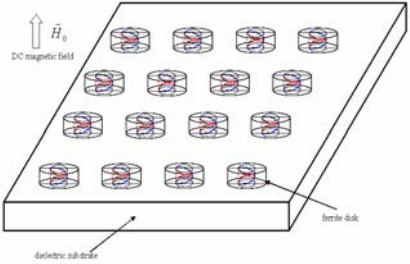

Fig. 6. A dense microwave ME composite material with an illustration of helical (chiral) states of the MDMs inside ferrite particles.

## 5. NOVEL COMPACT MICROWAVE RADIATING SYSTEMS

In open resonant microwave structures with ferrite inclusions there exist the vortex-type fields and Poynting-vector phase singularities. A circularly polarized EM radiation obtained from a chiral-state vortex antenna with a small ferrite inclusion can be considered as being originated from a vortex topological defect in space. Recent studies of patch antennas with ferrite disk inclusions show unique microwave characteristics [16].

Fig. 7 shows the Poynting vector distributions above the patch antenna with a small ferrite-disk inclusion. At the same direction of a bias magnetic field, one has two chiral-state resonances at different frequencies.

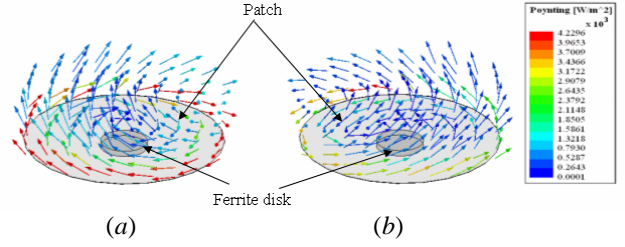

Fig. 7. The Poynting vector distributions above the patch for two resonance frequencies and the same direction of a bias magnetic field: (a) left-hand chiral resonant state; (b) right-hand chiral resonant state.

Novel compact microwave radiating systems can be realized when ferrite disks work at the MDM ME behavior.

## 6. MICROWAVE ME LOGIC DEVICES

Devices based on magnetic ordering may provide an interesting alternative to conventional semiconductor gates. MDM ME ferrite disks have well-spaced energy levels owing to there intrinsic spectral characteristics. They also have eigen electric moments. These properties make them feasible for electrostatic-control logic devices operating at room temperatures.

Presently, there is a very strong interest in realization of logic devices and computation systems working based on the quantum-type algorithms. Since MDMs are energetically orthogonal, there is no magnetic dipole-dipole interaction between the spins of different modes. This allows considering MDMs as Hilbert-space oscillations. Recent propositions [17], [18] show unique perspective in use of ferrite ME particles for realization of novel room-temperature quantum-type computation devices. One of the schemes can be created with use of two coupled MDM ME particles as a quantum-type gate. Another implementation is based on the adiabatic transfer of ME-particle-state coherence to the cavity mode. The initial eigenstates describe the ME-particle-cavity system. The final eigenstates contains a contribution from the excited ME-particle state.

Let $f_1$ be a resonance frequency of cavity mode $TE_{101}$ and $f_2$ be a resonance frequency of cavity mode $TE_{201}$. It is clear that a ME particle placed in a cavity center may interact with the cavity electric field in a case of $TE_{101}$ mode and with the cavity magnetic field in a case of $TE_{201}$ mode. The high-$Q$ microwave cavity with these two resonance frequencies can be viewed, respectively, as a two-state system: $|0\rangle$ and $|1\rangle$. The particle characteristic dimension is much less than the EM-wave wavelength (in Fig. 8 we premeditatedly increased the particle sizes for clearer observation). The main absorption peak (the first-order ME mode) corresponds to the "ground" $|g\rangle$ state of a ME particle and the second absorption peak (the second-order ME mode) is the "excited" $|e\rangle$ state. Suppose that we realized a cavity with frequencies $f_1$ and $f_2$ corresponding to frequencies of the ME-particle first two peaks at DC magnetic $H_0^{(I)}$. The initial state $|\psi_{initial}\rangle = |g,0\rangle$ describing the ME-particle-cavity system corresponds to the case of the main ME mode at frequency $f_1$, bias magnetic field $H_0^{(I)}$ and cavity mode $TE_{101}$ (Fig. 9 (a)). For a certain magnetic field $H_0^{(II)} < H_0^{(I)}$ we can shift the oscillating spectrum of the particle to the position shown in Fig. 9 (b). In this case a RF electric field of cavity mode $TE_{101}$ excites the second-mode ME oscillation in a particle. This is the $|e,0\rangle$ state. Now we turn a DC magnetic field back to quantity $H_0^{(I)}$ (Fig. 9 (c)). The final cavity eigenstate ($TE_{201}$ mode) contains a quantized contribution from the excited ME-particle state. As a result, we have a "shift" through the transformations $|g,0\rangle \rightarrow |e,0\rangle \rightarrow |g,1\rangle$.

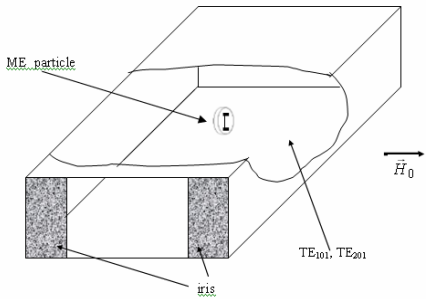

Fig. 8. ME particle inside a high-$Q$ cavity: The quantum-type entanglement between the external (cavity mode) and the internal (MDM) states.

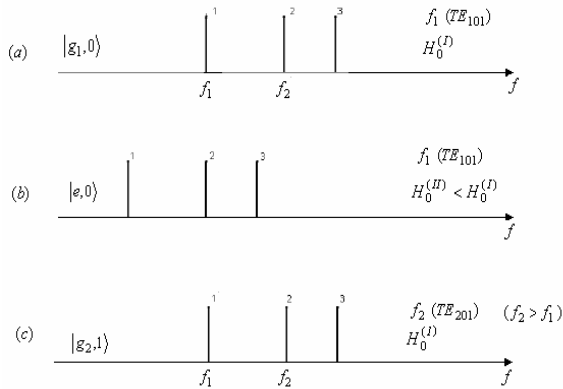

Fig. 9. A scheme illustrating how coherence of the MDM levels is mapped directly onto a cavity field.

The above mechanism shows how the source (the control qubit – the ME particle) can "teach" the field to evolve toward a desirable quantum state.

## 7. CONCLUSION

We showed that novel microwave devices with unique characteristics for both near-field and far-field manipulations can be realized based on ferrite ME particles.